\documentclass[twocolumn,aps,pra,showpacs,superscriptaddress,nofootinbib]{revtex4-1}

\usepackage[utf8]{inputenc} 
\usepackage[T1]{fontenc} 
\usepackage[USenglish]{babel}
\usepackage[sc,osf]{mathpazo}
\usepackage[scaled=0.86]{berasans} 
\usepackage[scaled=1.03]{inconsolata} 
\usepackage{amsmath,amssymb,amsthm,mathtools} 
\usepackage[colorlinks]{hyperref} 
\usepackage{graphicx}
\usepackage{xspace}

\DeclareMathOperator{\erf}{erf}

\let\originalleft\left
\let\originalright\right
\renewcommand{\left}{\mathopen{}\mathclose\bgroup\originalleft}
\renewcommand{\right}{\aftergroup\egroup\originalright}

\newcommand{\bra}[1]{\left\langle #1 \right|}
\newcommand{\ket}[1]{\left| #1 \right\rangle}
\newcommand{\braket}[2]{\left\langle #1 \middle| #2 \right\rangle}
\newcommand{\ketbra}[2]{\left|#1\middle\rangle\middle\langle#2\right|}
\newcommand{\exval}[3]{\left\langle #1 \middle| #2 \middle| #3 \right\rangle}

\newcommand{\abs}[1]{\left|#1\right|}
\newcommand{\mean}[1]{\left\langle#1\right\rangle}

\newcommand{\de}[1]{\left(#1\right)}

\newcommand{\DE}[1]{\left\{#1\right\}}

\newcommand{\chsh}{\mathcal{B}}

\newcommand{\id}{\mathbb{1}}

\DeclareMathOperator{\dint}{d\!}

\newcommand{\ie}{\emph{i.e.}\@\xspace}
\newcommand{\etal}{\emph{et al.}\@\xspace}

\mathtoolsset{centercolon}

\hypersetup{pdftitle={{Tests of Bell inequality with arbitrarily low photodetection efficiency and homodyne measurements}}}

\begin{document}

\title{Tests of Bell inequality with arbitrarily low photodetection efficiency and homodyne measurements}


\author{Mateus Araújo}
 \affiliation{Departamento de Física, Universidade Federal de Minas Gerais,
 Caixa Postal 702, 30123-970 Belo Horizonte, MG, Brazil}

\author{Marco Túlio Quintino}
 \affiliation{Departamento de Física, Universidade Federal de Minas Gerais,
 Caixa Postal 702, 30123-970 Belo Horizonte, MG, Brazil}

\author{Daniel Cavalcanti}
 \affiliation{Centre for Quantum Technologies, National university of Singapore, 3 Science drive, Singapore 117543}

\author{Marcelo França Santos}
 \affiliation{Departamento de Física, Universidade Federal de Minas Gerais,
 Caixa Postal 702, 30123-970 Belo Horizonte, MG, Brazil}

\author{Adán Cabello}
 \affiliation{Departamento de Física Aplicada II, Universidad de
 Sevilla, E-41012 Sevilla, Spain}
 \affiliation{Department of Physics, Stockholm University, S-10691
 Stockholm, Sweden}

\author{Marcelo Terra Cunha}
 \affiliation{Departamento de Matemática, Universidade Federal de Minas Gerais,
 Caixa Postal 702, 30123-970 Belo Horizonte, MG, Brazil}


\date{\today}



\begin{abstract} We show that hybrid local measurements combining homodyne measurements and photodetection provide violations of a Bell inequality with
\textit{arbitrarily low} photodetection efficiency. This is shown in two different scenarios: when one part receives an atom entangled to the field mode to be
measured by the other part and when both parts make similar photonic measurements. Our findings promote the hybrid measurement scenario as a candidate for
loophole-free Bell tests beyond previous expectations. \end{abstract}



\maketitle


\section{Introduction}


One of the greatest discoveries of modern science is that quantum mechanics predicts nonlocal correlations \cite{bell64}. Experiments have shown a good agreement
with theory \cite{FC72,ADR82,KMWZSS95,WJSWZ98,RKVSIMW01,MMMOM08,SUKRMHRFLJZ10}, but all of them have loopholes that prevent the definitive conclusion that nature
is intrinsically nonlocal and allows for secure communication \cite{BHK05, ABGMPS07} and genuine randomness \cite{PAMBMMOHLMM10}. The ``ultimate test of quantum
mechanics'' \cite{Merali11}, the loophole-free Bell experiment, is still pending.

Two requirements have never been simultaneously satisfied in a Bell test: spacelike separation between each observer's measurement choice and the other
observer's measurement outcome \cite{ADR82,WJSWZ98,SUKRMHRFLJZ10}, and overall detection efficiency $\eta$ (defined as the ratio between detected and emitted
particles) beyond a threshold value that rules out local models~\cite{RKVSIMW01,MMMOM08}. Spacelike separation is typically overcome by using photons as
information carriers. However, this makes it difficult to obtain $\eta$ higher than the threshold, as photodetection is usually inefficient.

For the Clauser-Horne-Shimony-Holt (CHSH) \cite{chsh69} inequality, three different thresholds are known, corresponding to three different scenarios:
(i) If we assume that all measurements are made with efficiency $\eta$, as is usual when, {\em e.g.}, the qubits are the polarization of photons, this
threshold is $\eta > 2/3$ \cite{larsson01}. (ii) If we assume that Alice's measurements are made with efficiency $\eta_A=1$ and Bob's with efficiency $\eta_B$, a condition that is close to what can be achieved in atom-photon experiments, the threshold is $\eta_B > 1/2$ \cite{CL07,BGSS07}. (iii) If both Alice and Bob measure one of their observables with efficiency $\eta_0=1$ and the other with efficiency $\eta_1$, the threshold is $\eta_1 > 0$ \cite{garbarino10}. Albeit very alluring, this last scenario lacks (until now) a physical system with which to be implemented.

	To overcome the detection loophole with photons, an alternative strategy is to use highly efficient homodyne measurements (quadrature variables).
However, to date, only small Bell violations were predicted using feasible states and homodyne measurements \cite{grangier04,nhacar04}. Recently, Ji \etal
\cite{JKJZN10} (see also \cite{cavalcanti11}) introduced a promising idea: they proposed combining homodyne measurements (quadrature variables) of high quantum
efficiency with photon number measurements.

It has been shown that this new tool indeed allows for Bell tests requiring lower threshold photodetection efficiencies in the photon-photon scenario: $\eta >
0.71$ with feasible states \cite{CBSSV11} and even $\eta > 0.29$ for more general states \cite{QACST11}. The atom-photon case has also been recently discussed
\cite{SBGRSWW11}. However, in practical terms, the advantage when comparing to previous proposals is, so far, too small to stimulate Bell experiments exploiting
hybrid local measurements.

In this paper, we show that the hybrid measurement scenario can be used to implement a variant of the scheme proposed by Garbarino \cite{garbarino10}, and
therefore allows a violation of the CHSH inequality even when the photodetectors of Alice and Bob have an arbitrarily low efficiency. To achieve this result, we
need to assume ideal homodyne detection and use some states which are difficult to prepare (see, however, \cite{teo12}). This finding  promotes hybrid local measurements as a candidate for photonic loophole-free Bell tests beyond previous expectations.

We discuss two experimental setups. In the first scenario, an entangled atom-light state is prepared. The electronic levels of the atom encode a qubit that can
be very efficiently detected \cite{HKHRWW10} and that remains with Alice, while the light field is sent to Bob who chooses between highly efficient homodyning
\cite{ourjoumtsev06} or inefficient photodetection. Bob's homodyne measurement is dichotomized in order to fit the CHSH inequality, as shall be explained later.
In this case, three out of the four possible measurements are very efficient and we calculate the threshold for Bob’s inefficient photodetection. In the second
setup, both Alice and Bob use photonic systems and perform homodyning (assumed to be nearly perfect) or photodetection (efficiency threshold to be calculated).


\section{CHSH observables}\label{sec:observables}


Consider the quantum operator corresponding to the CHSH inequality 
\begin{equation}
 \chsh:=A_0\otimes (B_0 + B_1) + A_1\otimes (B_0-B_1),
\end{equation} where $A_i$ and $B_j$ are observables with eigenvalues $\pm1$.

    When dealing with atomic systems we will consider that perfect Pauli measurements can be made. These measurements will be parametrized as
\begin{equation}
 V(\gamma) := \cos\gamma\ \sigma_z + \sin\gamma\ \sigma_x.
\end{equation}

	The treatment of the photonic observables will be a bit more involved, since they are the ones which are involved in the considerations of detection
efficiency. More specifically, we want to use them to implement the scheme from \cite{garbarino10} that achieves arbitrarily low detection efficiency.
However, this scheme is not applicable to the hybrid measurement scenario for two reasons: our quantum system -- the occupation number of photons -- is
infinite-dimensional, while the scheme is described for two qubits, and its definition of detection efficiency -- as the probability of detecting a particle --
is not applicable, since one of our measurements is precisely the presence or absence of a photon. A translation work is therefore required.

	To begin, we shall use the results of \cite{QACST11}, which showed that it is possible to reach the Tsirelson bound of $2\sqrt{2}$ in the hybrid measurement
scenario with states restricted to a two-qubit subspace. We shall see that precisely this same subspace allows the violation of the CHSH inequality with
arbitrarily low detection efficiency.

	Following \cite{QACST11}, we define our photodetection and homodyning observables as follows: the output of our photodetection observable is defined as $-1$
in the case of a click and $+1$ when there is no click. Photodetectors used for low intensity fields will be well represented, in the Fock basis, as
\begin{equation}
 D:=\ketbra{0}{0} - \sum_{n=1}^\infty \ketbra{n}{n}.
\end{equation} The quadrature measurement gives an infinite number of outcomes $x \in \mathbb{R}$. So, to use the CHSH inequality we need to dichotomize them, \ie, to use a binning process. The most general binning is that we output the value $+1$ if the $X$ measurement returns $x \in A^+$, where $A^+$ is a subset of
the real numbers. We output the value $-1$ if $x\in A^-=\mathbb{R}\setminus A^+$.

	Introducing the projectors \begin{equation}
 P_{Q\pm}:=\int_{A^\pm} \ketbra{x}{x}\dint x,
\end{equation} 	the dichotomized version of the $X$ measurement can be written as \begin{subequations} \begin{equation} 	Q = P_{Q^+} - P_{Q^-}, \end{equation} 	 
and its matrix representation in the Fock basis will be \begin{align}
 \exval{m}{Q}{n} &= \exval{m}{P_{Q+}}{n}-\exval{m}{P_{Q-}}{n}\nonumber \\
 &= 2\int_{A^+}\varphi_m(x)\varphi_n(x) \dint x- \delta_{mn},
\end{align} \end{subequations} where $\varphi_n(x)=\braket{x}{n}$ is the $n$th Hermite function.

	To find the adequate two-qubit subspace, first note that since $Q^2 = I$, the subspace spanned by $\DE{\ket{0}, Q\ket{0}}$ is invariant under $Q$, and we can therefore write \begin{subequations} \begin{equation}
 Q\ket{0} = \cos \theta \ket{0} + \sin \theta \ket{\Xi},
\end{equation} where \begin{equation}
 \ket{\Xi} := \frac{2}{\sin\theta}\sum_{n=1}^\infty \int_{A^+} \varphi_0(x) \varphi_n(x) \dint x\ket{n},
\end{equation} and \begin{equation}\label{eq:theta} \cos\theta := 2\int_{A^+}{\varphi_0(x)^2}\dint x-1. \end{equation} \end{subequations}
Since both $\ket{0}$ and $\ket{\Xi}$ are eigenvectors of $D$, the restriction of the observables $Q$ and $D$ to this subspace still has eigenvalues
$\pm1$, and therefore it allows us to reach the maximal violation $2\sqrt{2}$.

	Also note that these restricted observables can be written in the orthonormal basis $\{\ket{0},\ket{\Xi}\}$ in a simple way, namely \begin{subequations}
\begin{align}
 Q_R =& \cos\theta\ \sigma_z + \sin\theta\ \sigma_x,\\
 D_R =& \sigma_z.
\end{align} \end{subequations}

	Now we shall consider the effects of the main error sources in this set up. First, let us suppose that the photodetector has probability $\eta$ of detecting
a photon. In this case, it is usual to model the imperfect photodetection as an ideal detector preceded by an amplitude damping channel \cite{mandel95}. Thus we
write \begin{equation} D_\eta=\ket{0}\bra{0} + \sum_{k=1}^{\infty}
\Big[2(1-\eta)^k -1\Big]\ket{k}\bra{k}. \end{equation} Since we shall work with the restricted operators, we need to get the effects of $\eta$ on $D_R$. With the
aid of the function 
\begin{subequations} 
\begin{equation} 
H:[0,1]\to[0,1];\quad H(\eta) := (1-\exval{\Xi}{D_\eta}{\Xi})/2, 
\end{equation} 
we conveniently write
\begin{equation} D_{R\eta}=H(\eta)D_R + [1-H(\eta)]\id. 
\end{equation} 
\end{subequations}

Another source of errors is the transmittance $t \le 1$ which affects both measurements: photodetection and homodyning. It can also be modeled by an amplitude
damping channel. For the photodetection observable the transmittance and efficiency effects can be combined and we have \begin{equation} D_{R\eta t}=H(\eta t)D_R + [1-H(\eta t)]\id. \end{equation} Unfortunately, the homodyning observable with transmittance $t$ does not
admit such a simple representation.

	With this modeling of the detection efficiency, we now have effective observables that are formally equal to two-dimensional observables, and therefore the
analysis done in \cite{garbarino10} applies.


\section{Atom-photon scenario}


In this scenario Alice measures atomic observables, described as Pauli matrices, and Bob measures photodetection and X quadrature. We set the effective CHSH
operator as \begin{equation}
 \chsh_{A} := V(\gamma) \otimes (D_{\eta t} + Q_{t}) + V(-\gamma) \otimes (D_{\eta t} - Q_{t}),
\end{equation} where the observables were defined in section \ref{sec:observables}.

	As argued in the previous section we will restrict our attention to states belonging to the four-dimensional subspace spanned by
$\DE{\ket{g},\ket{s}}\otimes\DE{\ket{0},\ket{\Xi}}$, where $\ket{g}$ and $\ket{s}$ are two atomic levels. Also, we assume ideal transmittance ($t = 1$) and, for
simplicity (since it is sufficient), only consider binnings $A^+$ such that $\int_{A^+}{\varphi_0(x)^2}\dint x = 1/2$, where $Q_{R} = \sigma_x$.

	In this case, the Bell operator $\chsh_{AR}$ is represented by a $4\times 4$ matrix that can be easily diagonalized to find the eigenstate of the maximal
singular value \begin{multline}
 \ket{A_\Xi} \propto \\ \left[(1-H)\cos\gamma+\sqrt{\sin^2\gamma+(1-H)^2\cos^2\gamma}\right]\ket{g\,0}\\ + \sin\gamma\ket{s\,\Xi},
\end{multline} with the expectation \begin{equation}
 \exval{A_\Xi}{\chsh_{A}}{A_\Xi} = 2H\cos\gamma + 2\sqrt{\sin^2\gamma+(1-H)^2\cos^2\gamma}.
\end{equation} This is larger than $2$ if $\eta>0$ and $\gamma \in \de{0, \frac{\pi}{2}}$, as can be seen from simple algebra. That is, we have a violation of a
Bell inequality with \emph{arbitrarily low photodetection efficiency} for any choice of noncommuting atomic observables.

	To understand the effects of transmittance, we set $\eta=1$ to find the critical value of $t$ above which violations can be found. We did a
numerical search, again restricting ourselves to states in the subspace spanned by $\DE{\ket{g},\ket{s}}\otimes\DE{\ket{0},\ket{\Xi}}$. We found out that the
critical transmittance depends on the binning choice, and the best value we found was for the binning $A^+ = [-\erf^{-1} 1/2,\erf^{-1} 1/2]$, where $\erf$ is the
error function. Then violations are obtained for $t\ge0.55$. 		Although the state $\ket{A_\Xi}$ does serve as a proof of principle for the violation of a
Bell inequality with arbitrarily low efficiency, it is not practical for a real experiment due to the unphysical nature of $\ket{\Xi}$: for instance, the mean
number of photons
 $\exval{\Xi}{N}{\Xi}$ diverges. To avoid this problem, we could just truncate $\ket{\Xi}$ to any finite dimension, and approximate its properties arbitrarily
well. However, this truncated state would still be very hard to produce in a real experiment.

\begin{figure}[ht] 		\centering 	\includegraphics{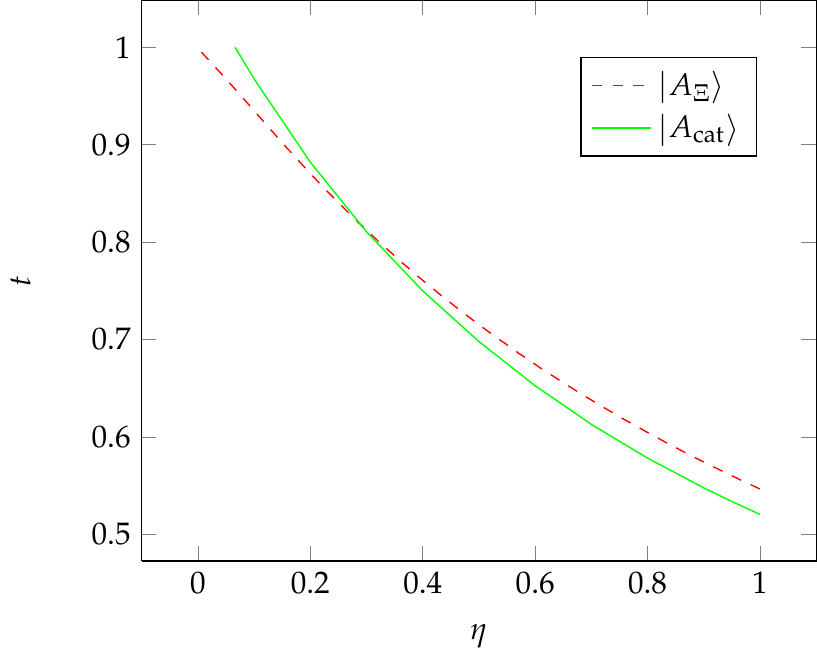} 	\caption{(Color online) Critical line on which $\mean{\chsh_A} = 2$, considering the efficiency $\eta$
of the photodetectors and the transmittance $t$ between the source and the photodetectors. For $\ket{A_\text{cat}}$ violations can be obtained above the critical efficiency $0.066$, for which $\alpha \approx 2.29i$, and above the critical transmittance $\approx0.52$, reached for $\alpha \approx 2.87i $. For $\ket{A_{\Xi}}$
the critical efficiency is zero, while the critical transmittance is $\approx 0.55$.} 	\label{fig:etat} 	\end{figure}

	It is physically sound to replace $\ket{\Xi}$ with a more familiar state. A good candidate is the so-called even cat state \cite{jeong05}: \begin{equation}
 \ket{\text{cat}} := \frac{\ket{\alpha}+\ket{-\alpha}}{\sqrt{2}\sqrt{1+ e^{-2\abs{\alpha}^2}}}.
\end{equation} The main reason behind this choice is that its fidelity with $\ket{\Xi}$ can be made very high. Doing so, the critical efficiency will no longer be arbitrarily low, but will still be very low. With this replacement, the state $\ket{A_\Xi}$ becomes \begin{equation}
 \label{with_cat}
 \ket{A_\text{cat}} = \cos\nu\ket{g\,0}+\sin\nu\ket{s\,\text{cat}},
\end{equation} where we shall optimize $\alpha$ and $\nu$ for each $\eta$ and $t$, since the functions defined in $\ket{A_\Xi}$ do not give the optimal result
for this state. Figure \ref{fig:etat} shows the boundary of the region of parameters $t$ and $\eta$ for which a CHSH violation is observed for these states. The
maximal CHSH value for the state \eqref{with_cat} is $2.60$, reached when $\alpha \approx 2.20i$, for $\eta = 1 = t$. 	

\section{Photon-photon scenario}


	In this scenario both Alice and Bob measure photodetection and $X$ quadrature. Now, the effective CHSH operator is
 \begin{equation}\label{chsh_P}
 \chsh_{P} := D_{\eta t} \otimes (D_{\eta t} + Q_t) + Q_t \otimes (D_{\eta t} - Q_t),
 \end{equation}
	with the observables defined in section \ref{sec:observables}. Note that now the family of $\chsh$ operators depends only on the binning choice, represented
by the parameter $\theta$, in contrast with the previous case where we could also optimize over the choice of the atomic observables, given by the $\gamma$
parameter.

\begin{figure}[ht] 		\centering 	\includegraphics{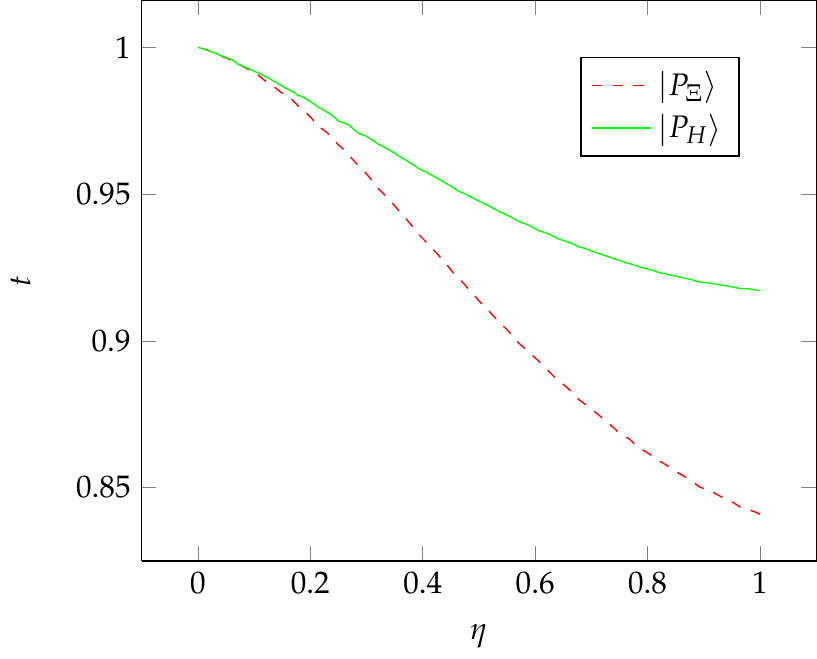} 	\caption{(Color online) Critical line on which $\mean{\chsh_P} = 2$, considering the efficiency
$\eta$ of the photodetectors and the transmittance $t$ between the source and the photodetectors.} 	\label{fig:etatfotonfoton} 	\end{figure}

	As before, let us restrict ourselves to a four-dimensional subspace, this time the one generated by $\{\ket{0},\ket{\Xi}\}^{\otimes 2}$, and call
$\mathcal{B}_{PR}$ the restriction of $\mathcal{B}_{P}$ to this subspace. By diagonalizing $\chsh_{PR}$ we can find a state $\ket{P_\Xi}$ that has a violation
for every $\eta>0$. This state has a complicated form and only provides a Bell violation for $t > 0.84$. In Fig. \ref{fig:etatfotonfoton}, we present the region
of parameters for which a violation of the CHSH inequality can be found using this state.


	On the other hand, irrespective of optimizations, we can take inspiration on Hardy's paradox to find a simple family of states that violates CHSH for all
$\eta>0$: 	\begin{equation} \ket{P_H}:=\frac{1}{\sqrt{1+\sin^2\frac{\theta}{2}}}\Big[{\cos\frac{\theta}{2} \ket{++} + \sin\frac{\theta}{2}
\de{\ket{+-}+\ket{-+}}}\Big], 	\end{equation} 	where $\ket{\pm}$ are the eigenstates of $Q_R$. The expectation value for $t=1$ is given by \begin{equation}
 \exval{P_H}{\chsh_{P}}{P_H}=2+ H^2\frac{4\cos^2\frac{\theta}{2}\sin^4\frac{\theta}{2}}{1+\sin^2\frac{\theta}{2}},
\end{equation} 	and is easily seen to be larger than $2$ for every $\eta>0$ and nontrivial choice of $\theta$. The critical transmittance of these states is
$\approx 0.92$, and is reached in the limit $\theta \to 0$. We have also maximized $\exval{P_H}{\chsh_{P}}{P_H}$ and found that the binning choice that reaches
the largest violation obeys $\cos\frac{\theta}{2}= (\sqrt{5}-1)/2$.


\section{Conclusions}


We have discussed atom-photon and photon-photon Bell tests combining homodyne measurements and photodetection. We have shown that, in both cases,
a Bell inequality can be violated even if the photodetectors used are arbitrarily inefficient, by assuming perfect detection
efficiency on the other observable.

	The photon-photon scenario considered here mimics, from the point of view of detection efficiencies, the scheme studied in \cite{garbarino10}. However, in
our schemes the detection efficiencies play a different role in the Bell test. While in \cite{garbarino10}, $\eta$ is related to the probability of a third
``no-click'' outcome, in our scenarios it is related to an attenuation process that cannot be directly noticed by the detectors.

	Under the present results, we see that highly efficient photodetectors are not required to achieve a loophole-free Bell test: indeed, the real problem is the
transmission. Note that schemes to circumvent this problem have been recently proposed \cite{gisin10,curty11,bohr12,cabello12}.
Also notice that under our scheme it is crucial that the propagating modes are very well matched to the modes to be detected. Although we do not present here an
experimental recipe to produce the quantum states considered (see \cite{teo12}), we hope that the present fidings can guide future research towards feasible
proposals within the present Bell scenario.


\begin{acknowledgments} This work was supported by the Brazilian agencies Fapemig, Capes, CNPq, and INCT-IQ, the National Research Foundation and the Ministry of
Education of Singapore, the Projects No.\ FIS2008-05596 and No.\ FIS2011-29400 (Spain), and the Wenner-Gren Foundation. D.C. is visiting UFMG under the program
PVE-CAPES (Brazil). \end{acknowledgments}


\bibliographystyle{linksen} 
\bibliography{biblio}

\end{document}